\documentclass[pra,showpacs,twocolumn,amsmath,a4paper]{revtex4}
\usepackage{bbm}
\usepackage{amssymb}
\usepackage{amsmath}
\usepackage{epsfig}
\usepackage{amsfonts}
\usepackage{latexsym}
\newcommand{\be}{\begin{equation}}
\newcommand{\ee}{\end{equation}}
\newcommand{\ba}{\begin{array}}
\newcommand{\ea}{\end{array}}
\newcommand{\bqa}{\begin{eqnarray}}
\newcommand{\eqa}{\end{eqnarray}}

\newcommand{\tr}{\mbox{Tr}}
\newcommand{\bra}[1]{\ensuremath{\langle #1 |}}
\newcommand{\ket}[1]{\ensuremath{| #1 \rangle}}

\input amssym.def
\begin{document}
\title{Measure of genuine multipartite entanglement with computable lower bounds}
\author{Zhi-Hao Ma$^{1}$, Zhi-Hua Chen$^{2}$, Jing-Ling Chen$^{3}$}
\email[]{ma9452316@gmail.com}
\email[]{chenjl@nankai.edu.cn}
\affiliation{$^{1}$Department of Mathematics, Shanghai Jiaotong
University, Shanghai, 200240, P.R.China\\
$^{2}$Department of Science, Zhijiang college, Zhejiang
University of technology, Hangzhou, 310024, P.R.China\\
$^{3}$Theoretical Physics
Division, Chern Institute of Mathematics, Nankai University,
Tianjin, 300071, P.R.China}

\author{Christoph Spengler, Andreas Gabriel, Marcus Huber}
\email[]{marcus.huber@univie.ac.at}

\affiliation{Faculty of Physics, University of Vienna, Boltzmanngasse 5, 1090 Vienna, Austria}

\begin{abstract}
We introduce an intuitive measure of genuine multipartite entanglement which is based on the well-known concurrence. We show how lower bounds on this measure can be derived that also meet important characteristics of an entanglement measure. These lower bounds are experimentally implementable in a feasible way enabling quantification of multipartite entanglement in a broad variety of cases.
\end{abstract}


\maketitle

\subsection{Introduction}
Entanglement is an essential component in quantum information and at the same time a
central feature of quantum mechanics \cite{Horodecki09,Guhne09}. Its potential applications in quantum information processing vary from quantum cryptography \cite{Ekert91} and quantum teleportation \cite{Bennett93} to measurement-based quantum computing \cite{BRaussendorf01}. The use of entanglement as a resource not only bears the question of how it can be detected, but also how it can be quantified. For this purpose, several entanglement measures have been introduced, one of the most prominent of which is the concurrence \cite{Wootters98,Horodecki09,Guhne09}. However, beyond bipartite qubit systems \cite{Wootters98} and highly symmetric bipartite qudit states such as isotropic states and Werner states \cite{Terhal00,Werner01} there exists no analytic method to compute the concurrence of arbitrary high-dimensional mixed states.
For a bipartite pure state $|\psi\rangle$ in a finite-dimensional Hilbert space $\mathcal{H}
_1\otimes \mathcal{H}_2=\mathbb{C}^{d_1}\otimes\mathbb{C}^{d_2}$ the concurrence is defined as \cite{Mintert05} $C(|\psi\rangle)=\sqrt{2\left(1-\texttt{Tr}\rho_1^2\right)}$ where
$\rho_1=\texttt{Tr}_2\rho$ is the reduced density matrix of $\rho=\ket{\psi}\bra{\psi}$. For mixed states $\rho$ the concurrence is generalized via the convex roof construction $C(\rho)=\inf_{\{p_i,|\psi_i\rangle\}} \sum_i p_i C(\ket{\psi_i})$ where the infimum is taken over all possible decompositions of $\rho$, i.e. $\rho=\sum_i p_i |\psi_i\rangle\langle\psi_i|$. This generalization is well-defined, however, as it involves a nontrivial optimization procedure it is not computable in general.
The concurrence is a useful measure with respect to a broad variety of tasks in quantum information which exploit entanglement between two parties. However, considering multipartite systems, a generalization of the concurrence is needed that strictly quantifies the amount of genuine multipartite entanglement - the type of entanglement that not only is the key resource of measurement-based quantum computing \cite{Briegel09} and high-precision metrology \cite{Giovannetti04} but also plays a central role in biological systems \cite{Sarovar,Caruso}, quantum phase transitions \cite{Oliv,Afshin} and quantum spin chains \cite{spinchains}.
Although many criteria detecting genuine multipartite entanglement have been introduced (see e.g. Refs.~\cite{Huber10,Huberqic,HuberDicke,Krammer,HuberClass,Deng09,Deng10,Chen10,Bancal,Horodeckicrit,Yucrit,Hassancrit,Seevinckcrit,Uffink,Collins,Guehnecrit}), there is still no computable measure quantifying the amount of genuine multipartite entanglement present in a system. There are only few quantities available for pure states (a set of possible measures is given in Ref.~\cite{HHK1}) which, however, are in general incomputable for mixed states and corresponding computable lower bounds have not been found so far.
In this paper, we define a generalized concurrence (analogously to a measure proposed for pure states in Ref.~\cite{Milburn}) for systems of arbitrarily many parties as an entanglement measure which distinguishes genuine multipartite entanglement from partial entanglement. As a main result we show that strong lower bounds on this measure can be derived by exploiting close analytic relations between this concurrence and recently introduced detection criteria for genuine multipartite entanglement.
\subsection{Genuine multipartite entanglement}
\begin{figure}[ht!]
\includegraphics[scale=0.20]{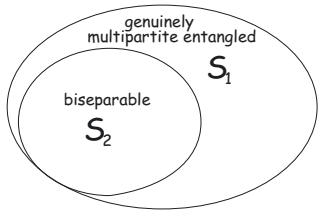}
 \caption{Illustration of the convex nested structure of multipartite entanglement. The set of biseparable states $\mathcal{S}_2$ is convexly embedded within the set $\mathcal{S}_1$ of all states ($\mathcal{S}_2 \subset \mathcal{S}_1$).}
  \label{fig_convex}
\end{figure}

An $n$-partite pure state $|\psi\rangle\in \mathcal{H}_1\otimes
\mathcal{H}_2\otimes\cdots\otimes\mathcal{H}_n$ is called biseparable if it can be written as
$|\psi\rangle=|\psi_A\rangle \otimes |\psi_B\rangle$, 
where $|\psi_A\rangle \in \mathcal{H}_{A} = \mathcal{H}_{j_1}\otimes \ldots \otimes \mathcal{H}_{j_k}$ and $|\psi_B\rangle \in \mathcal{H}_{B} = \mathcal{H}_{j_{k+1}}\otimes \ldots \otimes \mathcal{H}_{j_n}$ under any bipartition of the Hilbert space, i.e. a particular order $\{j_1,j_2,\ldots j_{k}|j_{k+1},\cdots
j_n \}$ of $\{1,2,\cdots, n\}$ (for example, for a 4-partite state, $\{1,3|2,4\}$ is a partition of $\{1,2,3,4\}$). An $n$-partite mixed state $\rho$ is
biseparable if it can be written as a convex combination of
biseparable pure states
 $\rho=\sum\limits_{i}p_i|\psi_i\rangle \langle\psi_i|$, 
wherein the contained $\{|\psi_i\rangle\}$ can be biseparable with respect to different
bipartitions (thus, a mixed biseparable state does not need to be separable w.r.t. any particular bipartition of the Hilbert space). If an $n$-partite state is not biseparable then it is
called genuinely $n$-partite entangled.\\
If we denote the set of all biseparable states by $\mathcal{S}_2$ and the set of all states by $\mathcal{S}_1$ we can illustrate the convex nested structure of multipartite entanglement (see Fig.~\ref{fig_convex}).\\
A measure of genuine multipartite entanglement (g.m.e.) $E(\rho)$ should at least satisfy:
\begin{itemize}
\item[M1] $E(\rho)=0 \,\forall\,\rho\in \mathcal{S}_2$ (zero for all biseparable states)
\item[M2] $E(\rho)>0 \,\forall\,\rho\in \mathcal{S}_1$ (detecting all g.m.e. states)
\item[M3] $E(\sum_ip_i\rho_i)\leq \sum_ip_iE(\rho_i)$ (convex)
\item[M4] $E(\Lambda_{LOCC}[\rho])\leq E(\rho)$ (non-increasing under local operations and classical communication)\footnote{On a single copy. The property of being non-increasing under LOCC in general cannot be required of any measure of genuine multipartite entanglement, as it was shown (e.g. in Ref.~\cite{Guhne09}) that genuine multipartite entanglement can locally be distilled out of a biseparable state if more copies are available.}
\item[M5] $E(U_{local}\rho U^\dagger_{local})= E(\rho)$ (invariant under local unitary transformations)
\end{itemize}
There are of course further possible conditions which are sometimes required (such as e.g. additivity), but this set of conditions constitutes the minimal requirement for any entanglement measure. For a more detailed analysis of such requirements consult e.g. Refs.~\cite{Mintertrep05,HHK1}. 
\subsection{Concurrence for genuine $n$-partite entanglement}
Let us now introduce a measure of multipartite entanglement satisfying all necessary conditions (M1-M5) for being a multipartite entanglement measure.\\
{\bf Definition 1.} For $n$-partite pure states $\ket{\Psi} \in \mathcal{H}_{1}\otimes \mathcal{H}_{2}\otimes\cdots\otimes \mathcal{H}_{n}$, where $dim(\mathcal{H}_{i})=d_{i},i=1,2, \cdots ,n$ we define the gme-concurrence as
\begin{align}
\label{gmeconcurrence}
 C_{gme}(\ket{\Psi}):=\min\limits_{\gamma_i \in \gamma} \sqrt{2(1-\tr(\rho^{2}_{A_{\gamma_i}}))}\ ,
\end{align}
where $\gamma=\{\gamma_i\}$ represents the set of all possible bipartitions $\{A_i|B_i\}$ of $\{1,2,\ldots,n\}$. The gme-concurrence can be generalized for mixed states $\rho$ via a convex roof construction, i.e.
\begin{align}
C_{gme}(\rho)= \inf_{\{p_i,|\psi_i\rangle\}} \sum_{i}p_{i}C_{gme}(\ket{\psi_i} ) \ ,
\label{4}
\end{align}
where the infimum is taken over all possible decompositions $\rho=\sum_i p_i \ket{\psi_i} \bra{\psi_i}$.
For example, for a tripartite pure state $\ket{\psi}\in \mathcal{H}_1 \otimes \mathcal{H}_2 \otimes \mathcal{H}_3$ there are three possible bipartitions $\gamma=\{ \{1|2, 3\}, \{2|1, 3\}, \{3|1, 2\} \}$. Consequently, the gme-concurrence is $C_{gme}(\psi)=\min\{\sqrt{2(1- Tr(\rho_{1}^2))},\sqrt{2(1- Tr(\rho_{2}^2))},\sqrt{2(1- Tr(\rho_{3}^2))}\}$.\\
The definition of $C_{gme}(\rho)$ directly implies $C_{gme}(\rho)=0$ for all biseparable states (M1) and $C_{gme}(\rho)>0$ for all genuinely $n$-partite entangled states (M2). Convexity (M3) also follows directly from the fact that any mixture $\lambda\rho_1+(1-\lambda)\rho_2$ of two density matrices $\rho_1$ and $\rho_2$ is at least decomposable into states that attain the individual infima. As the concurrence of any subsystem was proven to be non-increasing under LOCC (see e.g. Ref.~\cite{Mintert05}), the minimum of all possible concurrences will of course still remain non-increasing, thus proving (M4) also holds. Furthermore $\text{Tr}(\rho^2)$ is invariant under local unitary transformations for every reduced density matrix irrespective of the decomposition, which proves that also condition (M5) holds. For pure states the gme-concurrence is closely related to the entanglement of minimum bipartite entropy introduced for pure states in Ref.~\cite{Milburn}. In contrast to the original definition using von Neumann entropies of reduced density matrices, we use linear entropies. In this way we can derive lower bounds even on the convex roof extension which had not been considered before.
\subsection{Lower bounds on the gme-concurrence}
As the computation of any proper entanglement measure is in general an NP-hard problem (see Ref.~\cite{gurvits}), it is crucial for the quantification of entanglement that reliable lower bounds can be derived. These lower bounds should be computationally simple and also experimentally (locally) implementable to be of any use in practical applications. Let us now derive lower bounds on $C_{gme}$ which meet these requirements. Consider inequality II from Ref.~\cite{Huber10}, which is satisfied by all biseparable states (such that its violation implies genuine multipartite entanglement)
\begin{equation} \label{ineqII}
\underbrace{\sqrt{\bra{\Phi}\rho^{\otimes 2}\Pi_{\{1,2,\cdots,n\}}\ket{\Phi}}-\sum\limits_{\gamma}\sqrt{\bra{\Phi}\Pi_{\gamma}\rho^{\otimes 2}\Pi_{\gamma}\ket{\Phi}}}_{=: I[\rho,|\Phi\rangle]}\leq 0\, ,
\end{equation}
where $\ket{\Phi}$ is any state separable with respect to the two copy Hilbert spaces and $\Pi_{\{\alpha\}}$ is the cyclic permutation operator acting on the twofold copy Hilbert space in the subsystems defined by $\{\alpha\}$, i.e. exchanging the vectors of the subsystems $\{\alpha\}$ of the first copy with the vectors of the subsystems $\{\alpha\}$ of the second copy. A simple example would be 
$\Pi_{\{1\}}|\phi_1\phi_2\rangle\otimes|\psi_1\psi_2\rangle=|\psi_1\phi_2\rangle\otimes|\phi_1\psi_2\rangle$.\\
For sake of comprehensibility let us show how to derive lower bounds for three qubits and then generalize the result (in the appendix). If we consider a most general 3-qubit pure state in the computational basis
\begin{align}
\label{explicit}
|\psi\rangle=&&a|000\rangle+b|001\rangle+c|010\rangle+d|011\rangle\nonumber\\&&+e|100\rangle+f|101\rangle+g|110\rangle+h|111\rangle\, ,
\end{align}
the squared concurrences $C^2(\rho_\gamma)=2(1-\text{Tr}(\rho^2_\gamma))$ with respect to the three bipartitions read
\begin{align}
C^2(\rho_1)=4|ah-de|^2+F_1\,,\\
C^2(\rho_2)=4|ah-cf|^2+F_2\,,\\
C^2(\rho_3)=4|ah-bg|^2+F_3\,,
\end{align}
where $F_i$ are non-negative functions. The following relations thus hold
\begin{align}
C(\rho_1)\geq 2|ah-de|\geq2|ah|-2|de|\,,\\
C(\rho_2)\geq 2|ah-cf|\geq2|ah|-2|cf|\,,\\
C(\rho_3)\geq 2|ah-bg|\geq2|ah|-2|bg|\,,
\end{align}
and finally
\begin{align}
&\min\{C(\rho_1),C(\rho_2),C(\rho_3)\} \nonumber \\ \geq &2|ah|-2\max\{|de|,|cf|,|bg|\}\nonumber\\\geq&2|ah|-2(|de|+|cf|+|bg|)\ =: B \,.
\end{align}
Now for any given mixed state the convex roof construction is bounded by
\begin{align}
C_{gme}(\rho)\geq\inf_{\{p_i,|\psi_i\rangle\}}\sum_i p_iB_i\,.
\end{align}
For the choice $|\Phi\rangle=|000111\rangle$ and the abbreviation $\rho_{uvwxyz}:=\langle uvw|\rho|xyz\rangle$, inequality (\ref{ineqII}) reads
\begin{align}
I[\rho,|000111\rangle]=|\rho_{000111}|-\sqrt{\rho_{001001}\rho_{110110}}\nonumber\\-\sqrt{\rho_{010010}\rho_{101101}}-\sqrt{\rho_{100100}\rho_{011011}}\leq 0\,.
\end{align}
Now
\begin{align}
2|\rho_{000111}|\leq\inf_{\{p_i,|\psi_i\rangle\}}\sum_i p_i2|a_ih_i|\,,
\end{align}
due to the triangle inequality and
\begin{align}
2\sqrt{\rho_{001001}\rho_{110110}}\geq\inf_{\{p_i,|\psi_i\rangle\}}\sum_i p_i2|b_ig_i|\,,
\end{align}
holds due to the Cauchy-Schwarz inequality (of course for all the parts of the other bipartitions).\\
This leads to a lower bound on the convex roof construction
\begin{align}
C_{gme}(\rho)\geq 2I[\rho,|000111\rangle]\,.
\end{align}
As $C_{gme}(\rho)$ is invariant under local unitary transformations, we can infer that indeed every $2I[\rho,|\Phi\rangle]$ constitutes a proper lower bound. By taking into account the set of all vectors $\{|\Phi\rangle\}$ we can thus define a computable lower bound which itself has many favorable properties (satisfying M1, M3, M4 and M5):
\begin{align}
C_{gme}(\rho)\geq \max_{|\Phi\rangle}2I[\rho,|\Phi\rangle]\,.\label{lowerbound}
\end{align}
As the lower bound is straightforwardly generalized (the structure of the proof essentially remains the same, see the appendix for details), eq.(\ref{lowerbound}) is indeed a proper lower bound on (\ref{gmeconcurrence}) for any $n$-partite qudit state.\\
\begin{figure}[htp!]
\includegraphics[scale=0.30]{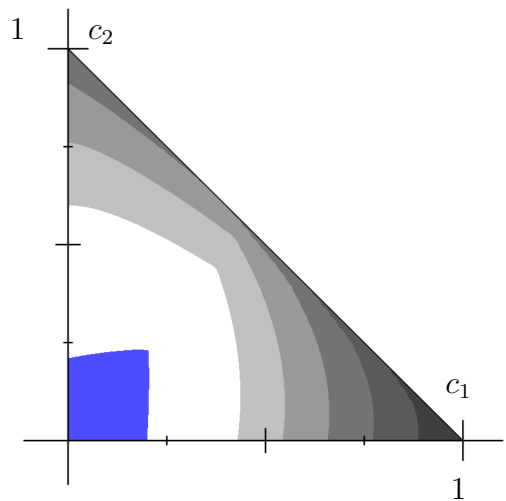}
\setlength{\unitlength}{1cm}%
  \begin{picture}(0, 0)(0,0)
\put(-0.8,0.2){\fontsize{12}{14}\selectfont \makebox(0,0)[]{$1$ \strut}}
\put(-6.6,6.3){\fontsize{12}{14}\selectfont \makebox(0,0)[]{$1$ \strut}}
\put(-0.8,1.6){\fontsize{12}{14}\selectfont \makebox(0,0)[]{$c_1$ \strut}}
\put(-5.5,6.3){\fontsize{12}{14}\selectfont \makebox(0,0)[]{$c_2$ \strut}}
\end{picture}
  \caption{Contour plot of the lower bound $\max_{|\Phi\rangle}2I[\rho,|\Phi\rangle]$ on the gme-concurrence for the set of three-qubit-states $\rho=c_1 \ket{GHZ}\bra{GHZ} + c_2 \ket{W}\bra{W}+\frac{1-c_1-c_2}{8}\mathbbm{1}$ given by convex mixtures of a GHZ state, W state and the maximally mixed state. The greyscale is related to the bound $\max_{|\Phi\rangle}2I[\rho,|\Phi\rangle]$ varying from $0$ to $1$ (where $0$ is white), while the blue region denotes states which are positive under partial transposition with respect to all bipartitions. The optimization over all $\{|\Phi\rangle\}$ was realized using the composite parametrization of the unitary group (see Ref.~\cite{SHH2}).}
  \label{fig_contourpl}
\end{figure}
\\
\textbf{Discussion}\\
The detection quality of our obtained bounds on the gme-concurrence is illustrated in Fig. \ref{fig_contourpl} for the family $\rho=c_1 \ket{GHZ}\bra{GHZ} + c_2 \ket{W}\bra{W}+\frac{1-c_1-c_2}{8}\mathbbm{1}$ of three-qubit-states, where
\begin{eqnarray} \ket{GHZ} = \frac{1}{\sqrt{2}}(\ket{000}+\ket{111}) \quad \textrm{and} \nonumber \\
\ket{W} = \frac{1}{\sqrt{3}}( \ket{001} + \ket{010} + \ket{100}) \end{eqnarray}
are the well-known genuinely multipartite entangled $GHZ$- and $W$-state, respectively. It can be seen that the bounds are non-zero for a considerable amount of multipartite entangled states, especially in the vicinity of the GHZ-state.\\
In fact, our bounds are exact for GHZ-like states, i.e. states of the form $\ket{gGHZ}=\alpha \ket{0'}^{\otimes n}+\beta \ket{1'}^{\otimes n}$ wherein $\ket{0'}\in\mathcal{H}_i$ and $\ket{1'}\in\mathcal{H}_i$ are arbitrary mutually orthogonal vectors. By expanding $\ket{gGHZ}$ in terms of $\ket{0'}$ and $\ket{1'}$ analogously to (\ref{explicit}) one finds $C(\rho_{A_{\gamma_i}})=2|\alpha \beta|\,\forall\,\gamma_i$, hence $C_{gme}(\rho)=2|\alpha \beta|$. In order to prove the exactness of the bound we choose $\ket{\phi}=\ket{0'}^{\otimes n}\ket{1'}^{\otimes n}$ for inequality II which then yields $2I\left[\ket{gGHZ}\bra{gGHZ},\ket{\phi}\right]=2|\alpha \beta|$. In fact we already know from the results of Ref.~\cite{Huber10}, that the inequality will detect a huge amount of genuinely multipartite entangled mixed states in arbitrary high dimensional and multipartite systems. In all of these situations we thus also have a lower bound on the gme-concurrence.\\
\\
\textbf{Experimental Implementation}\\
In order to be useful in practice, measures for multipartite entanglement need to be experimentally implementable by means of local observables (since all particles of composite quantum systems may not be available for combined measurements) without resorting to a full quantum state tomography (since the latter requires a vast number of measurements, which is unfeasible in practice). The bound (\ref{lowerbound}) satisfies these demands, as for fixed $\ket{\Phi}$ its computation only requires at most the square root of the number of measurements needed for a full state tomography. Furthermore it can be implemented locally as explicitly shown in \cite{Huberqic}. In an experimental situation where one aims at producing a certain state $\ket{\psi}$, it is now possible to choose the corresponding $\ket{\phi}$ and not only detect the state as being genuinely multipartite entangled, but also have a reliable statement about the amount of multipartite entanglement the state exhibits. Even if the produced state deviates from the desired states, the criteria are astonishingly noise robust (as e.g. analyzed in Ref.~\cite{Huber10}), as for example a GHZ state mixed with white noise is shown to be genuinely multipartite entangled with a white noise resistance of $\approx57\%$.\\
\\
\textbf{Conclusion}\\
We introduced a measure of genuine multipartite entanglement, which can be lower bounded by means of one of the currently most powerful detection criteria. These bounds are experimentally implementable and computationally very efficient, allowing to not only detect, but also to quantify genuine multipartite entanglement in an experimental scenario. This has grave implications on applications where genuine multipartite entanglement is a crucial resource (as e.g. in quantum computing \cite{BRaussendorf01} or cryptopgraphy \cite{SHH3}) and might allow to give a good estimate of the relevance of genuine multipartite entanglement in other physical systems (as e.g. in biological systems \cite{Caruso} or quantum spin chains \cite{spinchains}).\\
\\
{\bf Acknowledgement}: A. Gabriel, M. Huber and Ch. Spengler gratefully acknowledge the support of the Austrian FWF (Project P21947N16). J.L. Chen is supported by NSF of China (Grant No.10975075).
Z.H. Ma is supported by NSF of China(10901103), partially supported by a grant of science
and technology commission of Shanghai Municipality (STCSM, No.09XD1402500).\\
\\
\textbf{Appendix}\\
Let us finish by proving the lower bound for the general n-qudit case. For the most general pure n-qudit state $|\psi\rangle=\sum_{i_1,i_2,(\cdots),i_n}c_{i_1,i_2,(\cdots),i_n}|i_1i_2(\cdots)i_n\rangle$
the squared concurrences $C^2(\rho_\gamma)=2(1-\text{Tr}(\rho^2_\gamma))$ with respect arbitrary bipartitions ($\gamma$) always take the form
\begin{align}
C^2(\rho_\gamma)=4|c_{00(\cdots)0}c_{11(\cdots)1}-c_{\alpha(\gamma)}c_{\beta(\gamma)}|^2+F_\gamma\,,\\
\end{align}
where $F_\gamma$ are non-negative functions (see e.g. Refs.~\cite{HH2,HHK1} for details on how to calculate the linear entropies of arbitrary subsystems). For every bipartition $\gamma$ there exists one pair $\alpha(\gamma)$ and $\beta(\gamma)$ that can be retrieved from
\begin{align}
\{\alpha(\gamma),\beta(\gamma)\}=\pi_\gamma\{00(\cdots)0,11(\cdots)1\}\,
\end{align}
where $\pi_\gamma$ permutes every number from the subset defined by $\gamma$ from the first half of the joint set with the second. Thus
\begin{align}
C(\rho_\gamma)\geq 2|c_{00(\cdots)0}c_{11(\cdots)1}-c_{\alpha(\gamma)}c_{\beta(\gamma)}|\,
\end{align}
will hold also for every $\gamma$. Now for the GME-concurrence we can infer
\begin{align}
&\min_\gamma\{C(\rho_\gamma)\} \geq 2|c_{00(\cdots)0}c_{11(\cdots)1}|-(\sum_\gamma|c_{\alpha(\gamma)}c_{\beta(\gamma)}|)\, =: B \,.
\end{align}
Now for any given mixed state the convex roof construction is bounded by
\begin{align}
C_{gme}(\rho)\geq\inf_{\{p_i,|\psi_i\rangle\}}\sum_i p_iB_i\,.
\end{align}
For the choice $|\Phi\rangle=|0\rangle^{\otimes n}\otimes|1\rangle^{\otimes n}$, inequality (\ref{ineqII}) reads
\begin{align}
I[\rho,|\Phi\rangle]=|\rho_{00(\cdots)011(\cdots)1}|-\sum_\gamma\sqrt{\rho_{\alpha(\gamma)\alpha(\gamma)}\rho_{\beta(\gamma)\beta(\gamma)}}\leq 0\,.
\end{align}
Now
\begin{align}
2|\rho_{00(\cdots)011(\cdots)1}|\leq\inf_{\{p_i,|\psi_i\rangle\}}\sum_i p_i2|c^i_{00(\cdots)0}c^i_{11(\cdots)1}|\,,
\end{align}
due to the triangle inequality and
\begin{align}
\sqrt{\rho_{\alpha(\gamma)\alpha(\gamma)}\rho_{\beta(\gamma)\beta(\gamma)}}\geq\inf_{\{p_i,|\psi_i\rangle\}}\sum_i p_i2|c_{\alpha(\gamma)}c_{\beta(\gamma)}|\,,\\
\end{align}
due to the Cauchy-Schwarz inequality.\\
This leads to a lower bound on the convex roof construction
\begin{align}
C_{gme}(\rho)\geq 2I[\rho,|\Phi\rangle]\,.
\end{align}
And again due to the local unitary invariance of $C_{gme}(\rho)$ this proves our lower bound for all $|\Phi\rangle$.


\begin{thebibliography}{99}

\bibitem{Horodecki09} R. Horodecki, P. Horodecki, M. Horodecki, K. Horodecki,  Rev. Mod. Phys. \textbf{81}, 865 (2009).

\bibitem{Guhne09} O. G\"uhne, G. Toth, Phys. Rep. \textbf{474}, 1 (2009).

\bibitem{Ekert91}A. K. Ekert, Phys. Rev. Lett.  \textbf{67}, 661 (1991).

\bibitem{Bennett93} C. H. Bennett, G. Brassard, C. Cr¡äepeau, R. Jozsa, A. Peres, and W. K. Wootters, Phys. Rev. Lett.
 \textbf{70}, 1895 (1993).

\bibitem{BRaussendorf01} R. Raussendorf and H. J. Briegel, Phys. Rev. Lett.  \textbf{86}, 5188 (2001).

\bibitem{Wootters98} W. K. Wootters, Phys. Rev. Lett. {\bf 80}, 2245 (1998).

\bibitem{Terhal00}B.M. Terhal, K.H.G. Vollbrecht, Phys. Rev. Lett.\textbf{85}, 2625 (2000).

\bibitem{Werner01} K. G. H. Vollbrecht and R. F. Werner, Phys. Rev. A \textbf{64}, 062307 (2001).

\bibitem{Mintert05} F. Mintert, M. Kus, and A. Buchleitner, Phys. Rev. Lett. 95, 260502 (2005).

\bibitem{Briegel09} H. J. Briegel, D. E. Browne, W. D\"ur, R. Raussendorf and M. Van den Nest, Nat. Phys. \textbf{5}, 19 (2009).

\bibitem{Giovannetti04} V. Giovannetti, S. Lloyd and L. Maccone, Science  \textbf{306}, 1330 (2004).

\bibitem{Sarovar} M. Sarovar, A. Ishizaki, G.R. Fleming and K.B. Whaley, Nature Physics, \textbf{6}, 462 (2010).

\bibitem{Caruso} F. Caruso, A.W. Chin, A. Datta, S.F. Huelga and M.B. Plenio, Phys. Rev. A \textbf{81}, 062346 (2010).

\bibitem{Oliv} T.R. de Oliveira, G. Rigolin and M.C. de Oliveira, Phys. Rev. A {\bf 73}, 010305(R) (2006).

\bibitem{Afshin} A. Montakhab and A. Asadian, Phys. Review A \textbf{82}, 062313 (2010).

\bibitem{spinchains} D. Bru{\ss}, N. Datta, A. Ekert, L.C. Kwek and C. Macchiavello, 	Phys. Rev. A 72, 014301 (2005).

\bibitem{Huber10} M. Huber, F. Mintert, A. Gabriel and B. C. Hiesmayr, Phys. Rev. Lett. {\bf 104}, 210501(2010).

\bibitem{Huberqic} A. Gabriel, B.C. Hiesmayr, M. Huber, Quant. Inf. Comp.  \textbf{10}, 9\&10, 829(2010).
\bibitem{HuberDicke} M. Huber, P. Erker, H. Schimpf, A. Gabriel and B. C. Hiesmayr, Phys. Rev. A {\bf 83}, 022328(R) (2011).
\bibitem{Krammer} P. Krammer, H. Kampermann, D. Bru{\ss}, R. A. Bertlmann,
L. C. Kwek and C. Macchiavello, Phys. Rev. Lett. {\bf 103}, 100502 (2009).
\bibitem{HuberClass} M. Huber, H. Schimpf, A. Gabriel, Ch. Spengler, D. Bru{\ss}
and B.C. Hiesmayr, Phys. Rev. A {\bf 83}, 022328 (2011).
\bibitem{Bancal} J.-D. Bancal, N. Brunner, N. Gisin, and Y.-C. Liang, Phys. Rev. Lett. {\bf 106}, 020405 (2011).
\bibitem{Deng09} D. L. Deng, Z. S. Zhou, and J. L. Chen, Physical Review
A {\bf 80}, 022109 (2009).
\bibitem{Deng10} D. L. Deng, S. J. Gu, and J. L. Chen, Annals of Physics
{\bf 325}, 367372 (2010).
\bibitem{Chen10} J.-L. Chen, D.-L. Deng, H.-Y. Su, C. Wu, C. H. Oh, Phys.
Rev. A {\bf 83}, 022316 (2011).
\bibitem{Horodeckicrit} M. Horodecki, P. Horodecki and R. Horodecki, Phys.
Lett. A {\bf 283}, 1 (2001).
\bibitem{Yucrit} C.S. Yu and H.S. Song, Phys. Rev. A {\bf 72}, 022333 (2005).
\bibitem{Hassancrit} A. Hassan and P. Joag, Quantum Inf. Comput. {\bf 8}, 8\&9,
0773-0790 (2008).
\bibitem{Seevinckcrit} M. Seevinck and J. Uffink, Phys. Rev. A {\bf 78}, 032101
(2008).

\bibitem{Uffink}
J. Uffink, Phys. Rev. Lett. \textbf{88}, 230406 (2002).

\bibitem{Collins}
D. Collins, N. Gisin, S. Popescu, D. Roberts and V. Scarani, Phys. Rev. Lett. \textbf{88}, 170405 (2002).

\bibitem{Guehnecrit} O. G\"uhne and M. Seevinck, New J. Phys. {\bf 12}, 053002
(2010).

\bibitem{HHK1} B.C. Hiesmayr, M. Huber and Ph. Krammer, Phys. Rev.
A. {\bf 79}, 062308 (2009).
\bibitem{Milburn} D. T. Pope and G. J. Milburn, Phys. Rev. A {\bf 67}, 052107(2003).

\bibitem{Mintertrep05} F. Mintert, A. R.R. Carvalho, M. Kus, A. Buchleitner, Phys. Rep., {\bf 419}, 143 (2005).

\bibitem{gurvits} L. Gurvits, Proc. of the 35th ACM symp. on Theory of comp., pages 10-19, New York, ACM Press. (2003).

\bibitem{SHH2} Ch. Spengler, M. Huber and B.C. Hiesmayr, J. Phys. A: Math. Theor., {\bf 43}, 385306 (2010).

\bibitem{SHH3} S. Schauer, M. Huber and B.C. Hiesmayr, Phys. Rev. A {\bf 82}, 062311 (2010). 

\bibitem{HH2} B.C. Hiesmayr and M. Huber,	Phys. Rev. A {\bf 78}, 012342 (2008).

\end{thebibliography}
\end{document}